\documentclass[a4paper]{article}
\usepackage{amsmath}
\usepackage{amssymb}
\usepackage{graphicx}
\usepackage{wrapfig}
\usepackage{enumitem}
\usepackage{float}
\usepackage{geometry}
\usepackage{quantikz}
\usepackage{tabularx}
\usepackage{lscape}
\usepackage{rotating}
\usepackage{array}
\usepackage{caption}
\usepackage{wrapfig}
\usepackage{hyperref}
\usepackage{multirow}
\usepackage{makecell}
\usepackage{longtable}
\usepackage{subcaption}
\usepackage{hyperref}
\usepackage{bm}
\newcommand{\citep}[1]{\cite{#1}}

\date{}  
\title{Continuous-Variable Quantum Encoding Techniques: A Comparative Study of Embedding Techniques and Their Impact on Machine Learning Performance}

\begin{document}
	
\maketitle
\vspace{-1.5cm}
\author{\begin{center}
		Minati Rath\textsuperscript{*1} and Hema Date\textsuperscript{2}\\
		minati.rath.2019@iimmumbai.ac.in, hemadate@iimmumbai.ac.in\\
		\textsuperscript{1}\textsuperscript{, }\textsuperscript{2} Department of Analytics and Decision Science, Indian Institute of Management (IIM) Mumbai, India	
	\end{center}}

\begin{abstract}
This study explores the intersection of continuous-variable quantum computing (CVQC) and classical machine learning, focusing on CVQC data encoding techniques, including Displacement encoding and squeezing encoding, alongside Instantaneous Quantum Polynomial (IQP) encoding from discrete quantum computing.  We perform an extensive empirical analysis to assess the impact of these encoding methods on classical machine learning models, such as Logistic Regression, Support Vector Machines, K-Nearest Neighbors, and ensemble methods like Random Forest and LightGBM. Our findings indicate that CVQC-based encoding methods significantly enhance feature expressivity, resulting in improved classification accuracy and F1 scores, especially in high-dimensional and complex datasets. However, these improvements come with varying computational costs, which depend on the complexity of the encoding and the architecture of the machine learning models. Additionally, we examine the trade-off between quantum expressibility and classical learnability, offering valuable insights into the practical feasibility of incorporating these quantum encodings into real-world applications. This study contributes to the growing body of research on quantum-classical hybrid learning, emphasizing the role of CVQC in advancing quantum data representation and its integration into classical machine learning workflows.

\end{abstract}
\begin{keywords} \end{keywords}
\section{Introduction}

The rapid advancement of quantum computing has opened new frontiers in computational science, particularly in the realm of data representation and processing. Quantum technologies leverage fundamental principles such as superposition and entanglement, offering the potential to surpass classical computational limits \cite{feynman1982}\cite{Preskill2018}. A key area of interest in this paradigm shift is the encoding of classical data into quantum states—a process that directly influences the effectiveness of quantum-assisted machine learning models. While earlier studies have primarily focused on standard quantum encoding techniques such as basis encoding, angle encoding, and amplitude encoding \cite{huang2021power}, the exploration of more advanced quantum data representations remains an evolving and promising avenue of research.\\\\Machine learning, which has transformed industries ranging from healthcare to finance through data-driven decision-making \cite{miotto2018deep}\cite{rajpurkar2017chexnet} \cite{rath2023adaptivemodellingapproachrowtype}, stands to benefit significantly from enhanced data representations offered by quantum encoding techniques \cite{minati2025quantumpoweredcreditrisk}. Beyond discrete methods, continuous variable encoding strategies such as Instantaneous Quantum Polynomial (IQP) encoding, Displacement embedding, sqeezing embedding provide greater expressibility and richer feature mappings, potentially improving model performance in ways not achievable with classical approaches \cite{vcernotik2012displacement} \cite{douce2017continuous} \cite{yun2004squeezing} . These methods introduce structured transformations in the quantum Hilbert space, enabling new forms of feature representation that could refine classical machine learning models. However, the impact of these encoding strategies on machine learning performance, as well as their computational trade-offs, remains an open research question.\\\\This study aims to bridge this gap by conducting a comparative analysis of advanced quantum encoding methods and their influence on classical machine learning models. Unlike previous works that primarily examined conventional encoding techniques, we expand the scope to include next-generation quantum embeddings. Through rigorous experimentation across multiple machine learning algorithms—including Logistic Regression, Support Vector Machines, K-Nearest Neighbors, and ensemble methods such as Random Forest and LightGBM—we evaluate the extent to which these encodings enhance classification accuracy, F1 scores, and overall model efficiency. Additionally, we investigate the computational costs associated with each encoding method, providing insights into their practical feasibility for real-world applications. \\\\By systematically exploring these advanced quantum encoding strategies, this research contributes to the growing body of knowledge at the intersection of quantum computing and classical machine learning. Our findings offer valuable perspectives for researchers and practitioners seeking to harness quantum-inspired techniques for improved data analysis and predictive modeling. Furthermore, this study sheds light on the scalability and potential adoption of quantum encoding approaches in hybrid classical-quantum computational frameworks, paving the way for future advancements in quantum-enhanced machine learning.
\subsection{Quantum Computing}
Quantum computing represents a paradigm shift in information processing, leveraging fundamental principles of quantum mechanics to perform computations that are infeasible for classical systems. Unlike classical bits, which exist in discrete states of 0 or 1, quantum bits (qubits) can exist in a superposition of states, enabling simultaneous evaluation of multiple computational paths \cite{nielsen2000} \cite{rath2024quantumassistedsimulationframeworkdeveloping}. This inherent parallelism holds significant implications for enhancing machine learning models, particularly in tasks involving high-dimensional data representations.\\\\ Continuous-Variable Quantum Computing (CVQC) and Discrete-Variable Quantum Computing (DVQC) are two prominent paradigms in quantum computing \cite{choe2022quantum}. While DVQC typically uses qubits to represent information in binary states (0 or 1), CVQC encodes information in continuous variables, such as the amplitude and phase of light \cite{PhysRevLett.97.110501} \cite{menicucci2006universal} \cite{alexander2016flexible}. CVQC leverages quantum states of light, such as squeezed states and Gaussian states, to perform computations \cite{banchi2015quantum}. This continuous nature allows CVQC to efficiently handle certain types of data, particularly those involving real-valued inputs, which are common in machine learning applications.\\\\Entanglement, another cornerstone of quantum mechanics, establishes strong correlations between qubits regardless of spatial separation, facilitating more complex and efficient information encoding \cite{jaeger2009entanglement}. This property is particularly relevant for advanced quantum data encoding techniques, as it allows for richer feature mappings and interdependencies that classical representations struggle to capture. Quantum interference further refines computational outcomes by amplifying favorable solutions while canceling out less optimal ones, thereby optimizing the performance of quantum-enhanced algorithms.\\\\
Beyond these fundamental properties, quantum computing introduces novel approaches to information processing through specialized algorithms. Shor’s algorithm revolutionizes cryptographic security by factoring large numbers exponentially faster than classical methods \cite{shor1999polynomial}, while Grover’s algorithm enhances search processes with quadratic speedup \cite{grover1996fast}. In the context of quantum machine learning, Hamiltonian evolution and variational quantum circuits exploit quantum dynamics to encode and manipulate data in ways that classical models cannot \cite{cerezo2021variational}. These methods provide new avenues for feature transformation, particularly in hybrid quantum-classical frameworks. \\\\Despite its promise, practical quantum computing faces challenges such as decoherence, noise, and hardware limitations. Quantum error correction schemes are being developed to mitigate these issues, ensuring more reliable computation \cite{cerezo2021variational}. As quantum hardware matures, the integration of quantum-enhanced techniques with classical machine learning models is expected to drive significant advancements in computational efficiency and predictive performance. This study builds on these developments by exploring advanced quantum data encoding techniques and their implications for classical machine learning, aiming to uncover new strategies for optimizing data representation in hybrid quantum-classical environments.
\subsection{Quantum Data}

Quantum data refers to information encoded, stored, and processed using quantum bits (qubits), which operate according to the principles of quantum mechanics. Unlike classical data, which exists in binary states (0s and 1s), quantum data exploits superposition, entanglement, and interference to represent and process information more efficiently. These properties enable quantum data to encapsulate exponentially larger state spaces, facilitating more complex computations \cite{nielsen2000}.  \\\\ One of the defining characteristics of quantum data is superposition, which allows qubits to exist in a combination of states simultaneously. This property enhances computational efficiency by enabling parallel processing of multiple states within a single computation \cite{Preskill2018}. Additionally, entanglement establishes strong correlations between qubits, even when separated by vast distances, allowing for advanced applications in secure communication, quantum teleportation, and distributed quantum computing. Quantum interference further refines computation by amplifying desired outcomes while reducing the probability of incorrect results.\\\\ Quantum data manifests in multiple forms, each serving specific computational and cryptographic purposes:  

\begin{itemize}
	\item[] Quantum States: Quantum information is primarily represented as quantum states, which can range from basic qubit states $|0\rangle$ and $|1\rangle$ to more complex superposition and entangled states. These states serve as the foundation for quantum computing, quantum cryptography, and quantum simulation.  
	\item[] Quantum Registers: Collections of qubits or higher-dimensional qudits are organized into quantum registers, which function as the primary storage units in quantum algorithms and facilitate large-scale quantum computations.  
	\item[] Quantum Gates and Circuits: Quantum operations are implemented through quantum gates, such as the Hadamard gate (which introduces superposition), the CNOT gate (which entangles qubits), and the Toffoli gate (which implements controlled operations). These gates form quantum circuits, which execute quantum algorithms by manipulating qubits through sequences of unitary transformations \cite{Shor2002}.  
	\item[] Quantum Measurements: Unlike classical measurements, quantum measurements probabilistically collapse a quantum state into a definite classical outcome. This probabilistic nature introduces inherent uncertainty but also enables novel computational paradigms, such as quantum-enhanced optimization and cryptographic protocols.  
	\item[] Quantum Entanglement: Quantum data can encode complex correlations through entanglement, enabling applications like quantum key distribution (QKD), secure multi-party computation, and quantum-enhanced communication protocols \citep{gottesman2002introduction}.  
	\item[] Quantum Keys and Cryptographic Data: Quantum cryptography leverages quantum data to generate secure cryptographic keys using protocols such as BB84 \cite{bennett2014quantum} and E91 \cite{ekert1991quantum}. These quantum-generated keys provide security against eavesdropping due to the no-cloning theorem and quantum measurement principles \cite{brassard2000limitations}.  
\end{itemize}  

\noindent Additionally, emerging fields such as quantum image processing and quantum video encoding extend quantum data representations to visual and multimedia domains, potentially revolutionizing fields like quantum-enhanced artificial intelligence and secure quantum communication networks. \\\\ As quantum computing advances, efficient quantum data representation and encoding methods are becoming increasingly crucial. This study explores how quantum data encoding techniques can be leveraged to enhance classical machine learning models, improving feature transformations and enabling more expressive representations within hybrid quantum-classical frameworks.

\section{Literature Review}

The integration of quantum computing with classical machine learning has been an area of growing interest, offering potential advancements in data processing and computational efficiency \cite{rath2024quantum}. This literature review explores key developments in quantum data encoding techniques and their intersection with classical machine learning, focusing on foundational concepts, recent contributions, and prevailing challenges.\\\\Quantum data encoding plays a crucial role in bridging the gap between classical and quantum computing. Early contributions to this domain can be traced back to Feynman's seminal work on quantum simulations, which highlighted the potential of quantum systems to model complex physical phenomena more efficiently than classical counterparts \cite{feynman1982}. This laid the groundwork for subsequent research in quantum computing, particularly in quantum algorithms and encoding techniques.\\\\Preskill's introduction of Noisy Intermediate-Scale Quantum (NISQ) devices marked a significant milestone, emphasizing the feasibility of near-term quantum computers in solving computational problems \cite{Preskill2018}. This era of quantum computing has witnessed considerable advancements in quantum data encoding, enabling the application of quantum techniques to classical machine learning models.\\\\A growing body of research focuses on leveraging quantum-enhanced data representations to improve classical machine learning tasks. Rath et..al \cite{rath2024quantum} investigated how quantum data encoding could enhance traditional learning models, emphasizing the need to exploit quantum properties for improved feature transformations.\\\\Squeezed-state encoding has significant applications across various areas in quantum information science \cite{wu2024efficient}. One approach, explored by Li et al. (2021), uses squeezed-state encoding in quantum kernel methods to enhance quantum machine learning tasks by improving the feature space, which allows for more efficient data representation and processing \cite{li2022quantum}. Tiunov et al. (2024) focus on learning continuous-variable quantum states, utilizing squeezed light in multimode quantum state estimation, which helps in achieving more precise quantum predictions. In the context of quantum communication, Lau \& Plenio (2024) highlight the role of multimode squeezed states in quantum networks, reducing noise and improving communication fidelity over long distances \cite{PhysRevResearch.6.043113}. Finally, Yan et al. (2024) show the integral role of squeezed states in quantum teleportation, facilitating entanglement distribution and improving communication efficiency in quantum networks, thus enabling more robust and scalable quantum communication systems \cite{yan2024deterministic}. These advancements underscore the broad applicability of squeezed-state encoding in quantum technologies. \\\\ Despite these advancements, several challenges remain. Encoding classical data into quantum states is computationally expensive and constrained by the limited number of qubits available in current hardware \cite{wendin2017quantum}. Quantum error correction remains a significant hurdle, as decoherence and noise can distort encoded information \cite{nielsen2010quantum, devitt2013quantum}. Additionally, the interpretability of quantum machine learning models poses challenges, as quantum algorithms often function as black-box models.\\\\The practical implementation of quantum data encoding in real-world machine learning applications is still in its infancy. High computational resource demands and the limited accessibility of quantum hardware hinder broader adoption \cite{arute2019quantum, montanaro2016quantum}. While theoretical frameworks for quantum-enhanced learning are expanding, experimental validations and empirical studies are necessary to assess their real-world feasibility \cite{aaronson2016complexity, harrow2017quantum}.\\\\This study aims to contribute to this growing research area by empirically evaluating quantum data encoding techniques specifically tailored for tabular data in classical machine learning applications. By leveraging quantum simulation environments, we seek to provide insights into the comparative effectiveness of various encoding approaches, bridging the gap between theoretical advancements and practical implementations in hybrid quantum-classical learning frameworks.\\\\In summary, quantum data encoding presents promising opportunities for classical machine learning but also introduces significant computational and implementation challenges. This research endeavors to systematically investigate the impact of quantum encoding on classical models, shedding light on the potential benefits and limitations of this emerging paradigm.

\section{Implementation}
We implemented a set of machine learning algorithms using the PennyLane open-source software library to address a customer churn classification problem within the telecommunications domain.  The dataset consists of 20 features and a binary target column indicating churn, encompassing a total of 7,043 distinct customers. The dataset is publicly available on Kaggle. Among the features, categorical variables include gender, SeniorCitizen, Partner, Dependents, PhoneService, MultipleLines, InternetService, OnlineSecurity, OnlineBackup, DeviceProtection, TechSupport, StreamingTV, StreamingMovies, Contract, PaperlessBilling, PaymentMethod, and Churn. The numerical features are tenure, MonthlyCharges, and TotalCharges. Due to a high correlation coefficient (0.83) between tenure and TotalCharges, the latter was removed from the analysis. Additionally, PhoneService was eliminated due to high multicollinearity (VIF ~12), and MonthlyCharges was removed after reevaluating VIF scores (VIF ~6). The customerID feature was also excluded as it does not contribute to model learning.

\begin{wrapfigure}{r}{0.7\textwidth}
	\begin{center}
		\includegraphics[width=0.6\linewidth]{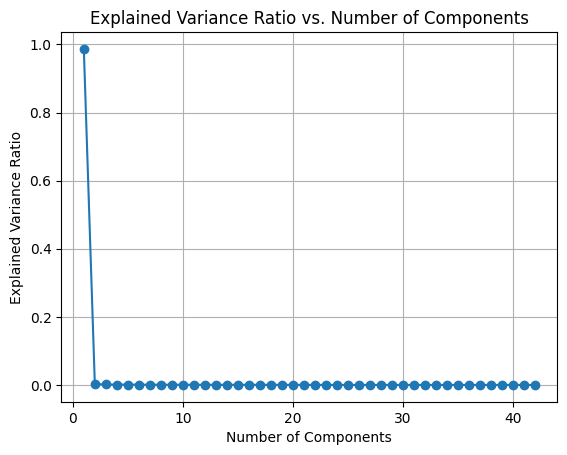}
		\caption{Explained Variance Ratio with Elbow Point}
		\label{fig:ExplainedVarianceElbow}
	\end{center}
	\begin{quote}
		\centering
		Elbow Point Index: 23\\
		Explained Variance Ratio at Elbow Point: 8.126234391114803e-33 \\
		Cumulative Explained Variance at Elbow Point: 1.0000000000000002 \\
	\end{quote}
\end{wrapfigure}

\noindent Following preprocessing, categorical features were one-hot encoded, resulting in a total of 42 columns. The dataset exhibited significant class imbalance, with 1,869 instances in the minority 'Yes' class and 5,174 in the majority 'No' class. Due to simulator limitations, we retained all minority class instances and performed undersampling on the majority class, yielding a balanced dataset of 3,738 records.\\\\To determine the optimal number of principal components, we plotted the explained variance ratio against the number of components. The elbow point, identified at 23 components, is shown in Figure \ref{fig:ExplainedVarianceElbow}. \\\\ For modeling, we employed various machine learning algorithms, including Logistic Regression, K-Nearest Neighbors (KNN), Support Vector Machines (SVM), and ensemble methods such as Random Forest, LightGBM, AdaBoost, and CatBoost. Logistic Regression was used as a baseline model due to its interpretability and effectiveness in binary classification. KNN was included for its instance-based learning approach, while SVM was selected for its ability to handle complex decision boundaries. The ensemble models were incorporated to leverage multiple weak learners and enhance predictive performance.\\\\Finally, quantum data encoding techniques were integrated into the workflow, employing basis encoding, angle encoding, and amplitude encoding. These methods facilitated the representation of classical data in quantum states, enabling the application of quantum computing principles to the churn prediction task. The quantum-enhanced models were evaluated in a hybrid quantum-classical pipeline, comparing their performance against classical counterparts.

\subsubsection{Instantaneous Quantum Polynomial (IQP) Encoding} 
Instantaneous Quantum Polynomial (IQP) encoding is a discrete Quantum encoding Method that utilizes a restricted form of quantum circuits with only commuting gates. IQP multi qubit  circuits are a non-universal model of quantum computation characterized by diagonal unitaries in the computational basis\cite{Havl_ek_2019}\cite{PhysRevA.111.012422} . It is particularly relevant in quantum machine learning and quantum supremacy experiments due to its unique structure that is believed to be classically hard to simulate.\\
\noindent IQP circuits consist only of gates that commute with each other (AB = BA), meaning their order does not affect the final state. These circuits follow the form:
\begin{equation}
	U_{\text{IQP}}(x) = H^{\otimes n} D(x) H^{\otimes n}
\end{equation}
where, where $H^{\otimes n}$ applies Hadamard gates to all qubits, and $D(\bm{x})$ is a diagonal unitary matrix encoding phase rotations. The diagonal unitary encoding is defined as:
\begin{equation} D(\bm{x}) = \exp\left(i(x_1 Z_1 + x_2 Z_2 + x_1 x_2 Z_1 Z_2)\right) \end{equation}
where $Z_i$ represents the Pauli-Z matrix:
$ Z = \begin{bmatrix} 1 & 0 \\ 0 & -1 \end{bmatrix} $. The Hadamard matrix is given by: $  H = \frac{1}{\sqrt{2}} \begin{bmatrix} 1 & 1 \\ 1 & -1 \end{bmatrix} $
and applying Hadamard gates to a two-qubit system results in:
\begin{equation}
	\begin{aligned}
		Z_1 &= Z \otimes I = \begin{bmatrix}
			1 & 0 & 0 & 0 \\
			0 & 1 & 0 & 0 \\
			0 & 0 & -1 & 0 \\
			0 & 0 & 0 & -1
		\end{bmatrix}, \quad
		Z_2 &= I \otimes Z = \begin{bmatrix}
			1 & 0 & 0 & 0 \\
			0 & -1 & 0 & 0 \\
			0 & 0 & 1 & 0 \\
			0 & 0 & 0 & -1
		\end{bmatrix}, \\
		Z_1 Z_2 &= Z \otimes Z = \begin{bmatrix}
			1 & 0 & 0 & 0 \\
			0 & -1 & 0 & 0 \\
			0 & 0 & -1 & 0 \\
			0 & 0 & 0 & 1
		\end{bmatrix}, \quad
		H \otimes H &= \frac{1}{2} \begin{bmatrix}
			1 & 1 & 1 & 1 \\
			1 & -1 & 1 & -1 \\
			1 & 1 & -1 & -1 \\
			1 & -1 & -1 & 1
		\end{bmatrix}
	\end{aligned}
\end{equation}

\noindent [$D(x)$] encodes the input data $x$ into the diagonal unitary. The key idea is that this encoding projects data into a highly non-linear quantum space, making it suitable for kernel-based quantum models. IQP circuits are believed to be BQP-complete, meaning they offer potential quantum advantage in machine learning. Unlike simple angle encoding, IQP encoding introduces non-linear correlations between features. IQP circuits form the basis for quantum kernel methods, improving classification tasks.\\\\ We begin by defining a classical input vector $\bm{x} = (0.5, 0.8)$, which we encode using an Instantaneous Quantum Polynomial (IQP) two qubit circuit. 
$x_1=0.5, x_2=0.8 $ \\
Step 1: Applying Equation 1 and 2. Compute the Matrix Sum =  $ 
0.5 Z_1 + 0.8 Z_2 + (0.5 * 0.8) Z_1 Z_2  $

\[ =
0.5
\begin{bmatrix}
	1 & 0 & 0 & 0 \\
	0 & 1 & 0 & 0 \\
	0 & 0 & -1 & 0 \\
	0 & 0 & 0 & -1
\end{bmatrix}
+
0.8
\begin{bmatrix}
	1 & 0 & 0 & 0 \\
	0 & -1 & 0 & 0 \\
	0 & 0 & 1 & 0 \\
	0 & 0 & 0 & -1
\end{bmatrix}
+
0.4
\begin{bmatrix}
	1 & 0 & 0 & 0 \\
	0 & -1 & 0 & 0 \\
	0 & 0 & -1 & 0 \\
	0 & 0 & 0 & 1
\end{bmatrix}
\]

Step 2: Compute Each Matrix Entry
\[
\begin{bmatrix}
	0.5 & 0 & 0 & 0 \\
	0 & 0.5 & 0 & 0 \\
	0 & 0 & -0.5 & 0 \\
	0 & 0 & 0 & -0.5
\end{bmatrix}
+
\begin{bmatrix}
	0.8 & 0 & 0 & 0 \\
	0 & -0.8 & 0 & 0 \\
	0 & 0 & 0.8 & 0 \\
	0 & 0 & 0 & -0.8
\end{bmatrix}
+
\begin{bmatrix}
	0.4 & 0 & 0 & 0 \\
	0 & -0.4 & 0 & 0 \\
	0 & 0 & -0.4 & 0 \\
	0 & 0 & 0 & 0.4
\end{bmatrix}
\]

Summing element-wise:
\[
M =
\begin{bmatrix}
	1.7 & 0 & 0 & 0 \\
	0 & -0.7 & 0 & 0 \\
	0 & 0 & -0.1 & 0 \\
	0 & 0 & 0 & -0.9
\end{bmatrix}
\]

Step 3: Apply the Exponential
Since \( D(x) = e^{iM} \), we exponentiate each diagonal element:

\[
D(x) =
\begin{bmatrix}
	e^{i(1.7)} & 0 & 0 & 0 \\
	0 & e^{i(-0.7)} & 0 & 0 \\
	0 & 0 & e^{i(-0.1)} & 0 \\
	0 & 0 & 0 & e^{i(-0.9)}
\end{bmatrix}
\]

\noindent These features allow quantum classifiers to separate data in a high-dimensional feature space efficiently.

\noindent The key insight is that IQP multi qubit encoding embeds classical data into quantum circuits via diagonal unitary transformations. This mapping enables quantum feature spaces where classical algorithms struggle to capture complex correlations. The resulting quantum state exhibits interference effects that enhance the expressiveness of features for machine learning applications.

\subsubsection{Displacement Encoding} 
Displacement Encoding is used to encode classical data into quantum states using displacement operators in phase space. It is particularly relevant when working with continuous-variable quantum computing (CVQC).  It allows for flexible transformations of quantum states, enhancing the expressiveness of quantum models.  The displacement operator is given by: $ D(\alpha) = e^{\alpha \hat{a}^\dagger - \alpha^* \hat{a}} $ 
where:
\begin{itemize}
	\item[] \(\alpha\) is the displacement amplitude (a complex number).
	\item[] \(\hat{a}^\dagger\) and \(\hat{a}\) are the creation and annihilation operators, respectively.  These operators are used to manipulate quantum states, such as the number (or Fock) states, which describe the quantum state of a system with a specific number of particles (like photons or quanta) \cite{cahill1969ordered} \cite{dirac1927quantum} \cite{wunsche1991displaced}. \\\\
	The annihilation operator \( \hat{a} \) lowers the number of particles (or quanta) in a quantum state by 1. For example, if the system is in a state with \( n \) quanta (denoted as \( |n\rangle \)), applying the annihilation operator \( a \) will give: $ a |n\rangle = \sqrt{n} |n-1\rangle $.  It "removes" one quantum (or particle) from the state, lowering the state by 1 quantum. \\	
	The creation operator \( \hat{a}^\dagger \) raises the number of particles (or quanta) in a quantum state by 1. 	Applying \( \hat{a}^\dagger \) to a state \( |n\rangle \) gives:
	$ \hat{a}^\dagger |n\rangle = \sqrt{n+1} |n+1\rangle $. It "adds" one quantum (or particle) to the state, increasing the state by 1 quantum.
\end{itemize}

\noindent For a two-qubit system, the displacement operator acts as: $
D(\alpha_1, \alpha_2) = e^{\alpha_1 a_1^\dagger - \alpha_1^* a_1} \otimes e^{\alpha_2 a_2^\dagger - \alpha_2^* a_2} $

\noindent For an \(N\)-qubit system, the displacement encoding is applied to each qubit independently. The combined displacement operator for the entire \(N\)-qubit system is: $ D(\alpha_1, \alpha_2, \dots, \alpha_N) = \bigotimes_{i=1}^{N} D_i(\alpha_i) $ where $ D_i(\alpha_i) = \exp\left( \alpha_i a_i^\dagger - \alpha_i^* a_i \right) $

\noindent After applying the displacement operators to all qubits, the resulting state will be: $ |\alpha_1, \alpha_2, \dots, \alpha_N\rangle = \bigotimes_{i=1}^{N} D_i(\alpha_i) |00 \dots 0\rangle $

\noindent To encode a classical value x=0.8 (Using single Qubit), \\
\indent Step 1: Displacement Operator The displacement operator is given by: $ 
D(\alpha) = e^{\alpha a^\dagger - \alpha^* a} $

For a real displacement \(\alpha = 0.8\) (imaginary part = 0, since 0.8 is real value),
 the equation \indent simplifies to: $ D(0.8) = e^{0.8 (a^\dagger - a)} $ 

Step 2: Computing \(|\alpha|^2\), Since \(\alpha\) is real: $  |\alpha|^2 = 0.8^2 = 0.64 $ 

Step 3: Poisson Distribution for Fock State Probabilities

The probability of being in the \(n\)-th Fock state in a coherent state \(|\alpha\rangle\) is: $ P(n) = \frac{|\alpha|^{2n}}{n!} e^{-|\alpha|^2} $

Substituting \(|\alpha|^2 = 0.64\):

For \(n = 0 \) (No photons present): $ 
P(0) = \frac{0.64^0}{0!} e^{-0.64} = e^{-0.64} = 0.5273  $

For \(n = 1\) (One photon present): $ P(1) = \frac{0.64^1}{1!} e^{-0.64} = 0.64 \times 0.5273 = 0.3375 $

For \(n = 2\) (Two photons present): $ P(2) = \frac{0.64^2}{2!} e^{-0.64} = \frac{0.4096}{2} \times 0.5273 = 0.1088 $

For \(n = 3\) (Three photons present): $ P(3) = \frac{0.64^3}{3!} e^{-0.64} = \frac{0.2621}{6} \times 0.5273 = 0.0190 $

For \(n = 4\) (Four photons present): $ P(4) = \frac{0.64^4}{4!} e^{-0.64} = \frac{0.1677}{24} \times 0.5273 = 0.0037 $ 

\begin{quote}
	\hangindent=1em
	\hangafter=0 Final Results:  For \(\alpha = 0.8\), the probabilities of finding the system in different Fock states are: $ P(0) = 0.5273, \quad P(1) = 0.3375, \quad P(2) = 0.1088, \quad P(3) = 0.0190, \quad P(4) = 0.0037 $.  The calculation for the displacement operator and the Poisson distribution is typically performed for the first few Fock states because the probability of higher n-values rapidly diminishes as  n increases. This is due to the nature of the Poisson distribution, where the probabilities for higher Fock states become increasingly small as n grows.
\end{quote}

\noindent The quantum state will look like: $ |\alpha = 0.8\rangle = P(0)|0\rangle + P(1)|1\rangle + P(2)|2\rangle + \cdots $, where the probabilities \( P(n) \) decrease as \( n \) increases.

\subsubsection{Squeezing Encoding} 

This method encodes classical data into squeezed vacuum states, exploiting the quantum mechanical property of squeezing to enhance precision in one quadrature at the expense of increased uncertainty in the conjugate quadrature.  Squeezed Comb State encoding utilizes squeezed comb states, which are finite superpositions of equidistant squeezed coherent states, to encode logical qubits. This method is designed to be robust against certain types of quantum noise. A single-mode squeezed vacuum state $|\zeta\rangle$ is obtained by applying the squeeze operator $S(\zeta)$ to the vacuum state $|0\rangle$: $ |\zeta\rangle = S(\zeta) |0\rangle $, where the squeeze operator is defined as: $
S(\zeta) = \exp\left(\frac{1}{2} \left( \zeta^* a^2 - \zeta (a^\dagger)^2 \right) \right) $  Where:  $a$ and $a^\dagger$ are the annihilation and creation operators and  $\zeta = r e^{i\phi}$ is the squeezing parameter, here $r$ is the squeezing amplitude and $\phi$ is the squeezing phase.

\noindent The quadrature operators: $ \hat{x} = \frac{a + a^\dagger}{\sqrt{2}}, \quad \hat{p} = \frac{a - a^\dagger}{i\sqrt{2}}
$ satisfy the Heisenberg uncertainty relation: $ \text{Var}(\hat{x}) \text{Var}(\hat{p}) \geq \frac{1}{4} $

\noindent In a squeezed state, the variances of the quadratures become:$
\text{Var}(\hat{x}) = \frac{e^{-2r}}{4}, \quad \text{Var}(\hat{p}) = \frac{e^{2r}}{4} $, which shows that squeezing reduces uncertainty in one quadrature at the cost of increasing uncertainty in the conjugate quadrature. \\\\ For an \( N \)-qubit system, the squeezing encoding is similarly applied to each qubit independently. The combined squeeze operator for the entire \( N \)-qubit system is: $ S(\zeta_1, \zeta_2, \dots, \zeta_N) = \bigotimes_{i=1}^{N} S_i(\zeta_i) $

where: $ S_i(\zeta_i) = \exp\left( \frac{1}{2} \left( \zeta_i^* a_i^2 - \zeta_i (a_i^\dagger)^2 \right) \right) $

After applying the squeeze operators to all qubits, the resulting state will be:

\[
|\zeta_1, \zeta_2, \dots, \zeta_N\rangle = \bigotimes_{i=1}^{N} S_i(\zeta_i) |00 \dots 0\rangle
\]

\noindent To encode the classical value \( x = 0.8 \) using squeezing encoding,\\\\Step 1: Choose the Squeezing Amplitude \( r \).  For the classical value \( x = 0.8 \), the squeezing amplitude \( r \) as a function of the value \( x \). So  \( r = 0.8 \), which will define the squeezing strength. \\
Step 2:  Quantum State After Squeezing the squeezed vacuum state \( |\zeta\rangle \) is obtained by applying the squeeze operator to the vacuum state \( |0\rangle \): $  |\zeta\rangle = S(\zeta) |0\rangle $ Where \( \zeta = r e^{i\phi} \) is the squeezing parameter, and for simplicity, let’s assume the squeezing phase \( \phi = 0 \) (real squeezing): $  S(0.8) = \exp\left( \frac{1}{2} \left( 0.8^* a^2 - 0.8 (a^\dagger)^2 \right) \right) $ \\
Step 3: The resulting state after applying the squeezing operator for the classical value \( x = 0.8 \) is: $ 
|\zeta\rangle = S(0.8) |0\rangle $ Where: $  S(0.8) = \exp\left( \frac{1}{2} \left( 0.8^* a^2 - 0.8 (a^\dagger)^2 \right) \right) $ \\
Step 4: The quadrature operators \( \hat{x} \) and \( \hat{p} \) are defined as: $  \hat{x} = \frac{a + a^\dagger}{\sqrt{2}}, \quad \hat{p} = \frac{a - a^\dagger}{i\sqrt{2}} $.  The uncertainty relation is: $ 
\text{Var}(\hat{x}) \text{Var}(\hat{p}) \geq \frac{1}{4} $.  In the squeezed state, the variances of the quadratures become: $  \text{Var}(\hat{x}) = \frac{e^{-2r}}{4}, \quad \text{Var}(\hat{p}) = \frac{e^{2r}}{4} $ 
For \( r = 0.8 \), these become: $  \text{Var}(\hat{x}) = \frac{e^{-1.6}}{4}, \quad \text{Var}(\hat{p}) = \frac{e^{1.6}}{4} $ \\
Step 5: The probability of finding the system in the \( n \)-th Fock state is given by the Poisson distribution: $ 
P(n) = \frac{|\alpha|^{2n}}{n!} e^{-|\alpha|^2} $.  For a coherent state, \( |\alpha|^2 = r^2 = 0.64 \), so: 
For \( n = 0 \) (no photons): $  P(0) = e^{-0.64} \approx 0.5273 $ \\
For \( n = 1 \): $ P(1) = 0.64 \times e^{-0.64} \approx 0.3375 $ \\
For \( n = 2 \): $ P(2) = \frac{0.64^2}{2!} e^{-0.64} \approx 0.1088 $  \\
For \( n = 3 \): $ P(3) = \frac{0.64^3}{3!} e^{-0.64} \approx 0.0190 $  \\
For \( n = 4 \): $ P(4) = \frac{0.64^4}{4!} e^{-0.64} \approx 0.0037 $ \\

\noindent Final Encoded State: The final quantum state after encoding the classical value \( x = 0.8 \) is represented by the squeezed vacuum state: $ |\zeta\rangle = S(0.8) |0\rangle$

\noindent This state encodes the classical information into the quantum system, where the squeezed quadratures exhibit reduced uncertainty in one direction (in this case, \( x \)-quadrature) at the cost of increased uncertainty in the conjugate quadrature (the \( p \)-quadrature).

\section{Experimental Results}
The performance of quantum data embedding in classical machine learning algorithms has been thoroughly investigated. The primary focus was on assessing the efficacy of advanced quantum-based data encoding techniques, including IQP, Displacement and Sqeezing, when applied to classical machine learning models. The classical algorithms employed in this study encompassed Logistic Regression, K-Nearest Neighbors (KNN), Support Vector Machine (SVM) with linear, polynomial, radial basis function (RBF), and sigmoid kernels, Decision Tree, Random Forest, LightGBM, AdaBoost, and CatBoost \ref{tab:Results}. The evaluation metrics such as accuracy, precision, sensitivity (recall), F1 score, and area under the receiver operating characteristic curve (ROC AUC) were used to comprehensively analyze and compare the performance of these algorithms with quantum-encoded data against classical encoding methods, specifically Principal Component Analysis (PCA). \\\\
In classification tasks, we used various evaluation metrics to assess the performance of machine learning models. The accuracy, defined as the ratio of correct predictions to the total number of predictions, provides an overall measure of classification correctness. Precision, or the positive predictive value, is the ratio of true positives to the sum of true positives and false positives, indicating the accuracy of positive predictions. Sensitivity, also known as recall or the true positive rate, measures the model's ability to capture all positive instances by dividing true positives by the sum of true positives and false negatives. The F1 score, a harmonic mean of precision and recall, balances the trade-off between these two metrics. Additionally, the Area Under the Receiver Operating Characteristic curve (ROC AUC) assesses the model's ability to discriminate between positive and negative instances across different probability thresholds. A higher ROC AUC signifies superior model performance, particularly useful when evaluating the discriminatory power of the model.The Cohen's Kappa coefficient is a statistic that measures the level of agreement between two raters beyond that which would be expected by chance\cite{cohen1960coefficient}.\\\\The table 1 below shows the performance of classical data and quantum data embedding with an 80:20 train-test split ratio.  \\\\ The experimental results indicate that different data encoding techniques, including Classical, IQP, Displacement, and Squeezing, exhibit varying levels of performance across multiple classifiers and PCA dimensions. The accuracy of Logistic Regression improves as the number of PCA components increases, reaching a peak of 79.01\% for PCA=23. Similarly, k-NN and SVM (Linear and Polynomial) show an increasing trend in performance with higher PCA components. \\\\IQP encoding maintains an accuracy of 49.33\% across all models and PCA settings. This suggests that the method either does not effectively capture the necessary data structure or introduces excessive noise that affects classification performance. In contrast, Displacement and Squeezing encodings, which fall under continuous-variable quantum encoding, demonstrate relatively better performance. Logistic Regression with Squeezing achieves a maximum accuracy of 73.26\% for PCA=23, showing an improvement over IQP. Similarly, k-NN with Displacement encoding attains an accuracy of 71.12\% for PCA=23. A similar trend is observed in SVM models, where Displacement and Squeezing encodings show marginal improvements with increasing PCA components.\\\\The impact of PCA components is evident, as increasing PCA dimensions enhances classification performance for all encoding techniques. PCA=23 consistently provides the best results, demonstrating the importance of feature dimensionality reduction in improving predictive accuracy. Despite these improvements, quantum encodings tend to have significantly higher running times. The computational cost is particularly high in models such as SVM with Displacement encoding, especially when using PCA=23. In contrast, Logistic Regression with IQP encoding exhibits low running times but does not deliver comparable accuracy, indicating that the method may require further optimization. \\\\The results suggest that different encoding strategies influence classification outcomes in distinct ways. While some quantum encodings show improvements with increasing PCA components, computational efficiency remains a key consideration. Further exploration is required to optimize these techniques for enhanced predictive performance and practical feasibility.

\begin{landscape}
	\begin{longtable}{|p{1.5cm}|p{1.0cm}|p{2.2cm}|p{1.5cm}|p{1.5cm}|p{1.7cm}|p{1.0cm}|p{1.0cm}|p{1.0cm}|p{1.3cm}|p{1.5cm}|}
		\caption{Classical data, Discrete and Continuous Variable Quantum Data Embedding Performance} 
		\label{tab:Results} \\
		\hline
		\textbf{Classifier} & \textbf{PCA} & \textbf{Encoding Type} & \textbf{Accuracy} & \textbf{Precision} & \textbf{Sensitivity} & \textbf{Recall} & \textbf{F1 Score} & \textbf{ROC AUC} & \textbf{Cohen's Kappa} & \textbf{Running Time} \\
		\hline
		\endfirsthead
		
		\multicolumn{11}{c}%
		{{\tablename\ \thetable{} -- continued from previous page}} \\
		\hline
		\textbf{Classifier} & \textbf{PCA} & \textbf{Encoding Type} & \textbf{Accuracy} & \textbf{Precision} & \textbf{Sensitivity} & \textbf{Recall} & \textbf{F1 Score} & \textbf{ROC AUC} & \textbf{Cohen's Kappa} & \textbf{Running Time} \\
		\hline
		\endhead
		
		\hline \multicolumn{11}{|r|}{{Continued on next page}} \\ \hline
		\endfoot
		
		\hline
		\endlastfoot
		
\multirow{12}{*}{\makecell[{{p{1.5cm}}}]{Logistic \\ Regression}}  & \multirow{4}{*}{\makecell[{{p{1.0cm}}}]{2}}
& Classical & 72.9947 & 0.7241 & 0.7546 & 0.7546 & 0.7390 & 0.7296 & 0.4595 & 0.0052 \\
&& IQP & 49.3316 & 0 & 0 & 0 & 0 & 0.5000 & 0 & 0.0020 \\
&& Displacement & 71.3904 & 0.7538 & 0.6464 & 0.6464 & 0.6960 & 0.7148 & 0.4288 & 0.0399 \\
&& Squeezing & 71.3904 & 0.7538 & 0.6464 & 0.6464 & 0.6960 & 0.7148 & 0.4288 & 0.0071 \\
\cline{2-11}
& \multirow{4}{*}{\makecell[{{p{1.0cm}}}]{15}} & Classical & 75.9358 & 0.7519 & 0.7836 & 0.7836 & 0.7674 & 0.7590 & 0.5184 & 0.0130 \\
&& IQP & 49.3316 & 0 & 0 & 0 & 0 & 0.5000 & 0 & 0.0185 \\
&& Displacement & 71.6578 & 0.7313 & 0.6966 & 0.6966 & 0.7135 & 0.7168 & 0.4334 & 0.0753 \\
&& Squeezing & 71.5241 & 0.7280 & 0.6992 & 0.6992 & 0.7133 & 0.7155 & 0.4307 & 0.0078 \\
\cline{2-11}
& \multirow{4}{*}{\makecell[{{p{1.0cm}}}]{23}} & Classical& 79.0106 & 0.7921 & 0.7941 & 0.7941 & 0.7930 & 0.7900 & 0.5801 & 0.0095 \\
&&IQP & 49.3316 & 0.0000 & 0.0000 & 0.0000 & 0.0000 & 0.5000 & 0.0000 & 0.0340 \\
&&Displacement & 72.5936 & 0.7254 & 0.7388 & 0.7388 & 0.7320 & 0.7258 & 0.4516 & 0.0676 \\
&&Squeezing & 73.2620 & 0.7361 & 0.7361 & 0.7361 & 0.7361 & 0.7326 & 0.4651 & 0.0129 \\
\hline
\multirow{12}{*}{\makecell[{{p{1.5cm}}}]{KNN}}  &
\multirow{4}{*}{\makecell[{{p{1.0cm}}}]{2}}
&Classical & 72.3262 & 0.7172 & 0.7493 & 0.7493 & 0.7329 & 0.7229 & 0.4461 & 0.0437 \\
&&IQP & 61.0963 & 0.5665 & 0.9894 & 0.9894 & 0.7205 & 0.6058 & 0.2138 & 0.0363 \\
&&Displacement & 65.5080 & 0.6090 & 0.8918 & 0.8918 & 0.7238 & 0.6519 & 0.3057 & 0.0668 \\
&&Squeezing & 65.5080 & 0.6090 & 0.8918 & 0.8918 & 0.7238 & 0.6519 & 0.3057 & 0.0436 \\
\cline{2-11}
& \multirow{4}{*}{\makecell[{{p{1.0cm}}}]{15}} 
&Classical & 69.9198 & 0.6766 & 0.7784 & 0.7784 & 0.7239 & 0.6981 & 0.3971 & 0.0995 \\
&&IQP & 66.0428 & 0.6640 & 0.6675 & 0.6675 & 0.6658 & 0.6603 & 0.3207 & 0.0440 \\
&&Displacement & 70.5882 & 0.6992 & 0.7361 & 0.7361 & 0.7172 & 0.7055 & 0.4112 & 0.0880 \\
&&Squeezing & 70.5882 & 0.6992 & 0.7361 & 0.7361 & 0.7172 & 0.7055 & 0.4112 & 0.0959 \\
\cline{2-11}
& \multirow{4}{*}{\makecell[{{p{1.0cm}}}]{23}} 
& Classical & 73.7968 & 0.7194 & 0.7916 & 0.7916 & 0.7538 & 0.7372 & 0.4751 & 0.0848 \\
&& IQP & 62.0321 & 0.6323 & 0.5989 & 0.5989 & 0.6152 & 0.6206 & 0.2410 & 0.2960 \\
&& Displacement & 71.1230 & 0.7139 & 0.7177 & 0.7177 & 0.7158 & 0.7111 & 0.4223 & 0.0995 \\
&& Squeezing & 70.7219 & 0.7073 & 0.7203 & 0.7203 & 0.7137 & 0.7070 & 0.4142 & 0.0154 \\
\hline 
\pagebreak
\multirow{12}{*}{\makecell[{{p{1.5cm}}}]{SVM \\ Linear}}  &
\multirow{4}{*}{\makecell[{{p{1.0cm}}}]{2}}
 & Classical & 73.1283 & 0.7247 & 0.7573 & 0.7573 & 0.7406 & 0.7309 & 0.4621 & 0.2453 \\
&& IQP & 49.3316 & 0 & 0 & 0 & 0 & 0.5 & 0 & 0.1057 \\
&& Displacement & 71.3904 & 0.7538 & 0.6464 & 0.6464 & 0.6960 & 0.7148 & 0.4288 & 0.3022 \\
&& Squeezing & 71.3904 & 0.7538 & 0.6464 & 0.6464 & 0.6960 & 0.7148 & 0.4288 & 0.1371 \\
\cline{2-11}
&\multirow{4}{*}{\makecell[{{p{1.0cm}}}]{15}} 
& Classical & 75.0000 & 0.7388 & 0.7836 & 0.7836 & 0.7606 & 0.7495 & 0.4995 & 0.5831 \\
&& IQP & 49.3316 & 0 & 0 & 0 & 0 & 0.5 & 0 & 0.1568 \\
&& Displacement & 71.3904 & 0.7538 & 0.6464 & 0.6464 & 0.6960 & 0.7148 & 0.4288 & 0.9139 \\
&& Squeezing & 71.3904 & 0.7538 & 0.6464 & 0.6464 & 0.6960 & 0.7148 & 0.4288 & 0.3331 \\
\cline{2-11}
& \multirow{4}{*}{\makecell[{{p{1.0cm}}}]{23}}
& Classical & 77.5401 & 0.7482 & 0.8391 & 0.8391 & 0.7910 & 0.7745 & 0.5500 & 0.6471 \\
&& IQP & 49.3316 & 0 & 0 & 0 & 0 & 0.5 & 0 & 0.3130 \\
&& Displacement & 71.3904 & 0.7538 & 0.6464 & 0.6464 & 0.6960 & 0.7148 & 0.4288 & 0.8031 \\
&& Squeezing & 71.3904 & 0.7538 & 0.6464 & 0.6464 & 0.6960 & 0.7148 & 0.4288 & 0.3902 \\
\hline
\multirow{12}{*}{\makecell[{{p{1.5cm}}}]{SVM Poly}}  &
\multirow{4}{*}{\makecell[{{p{1.0cm}}}]{2}}
 & Classical & 71.7914 & 0.6714 & 0.8681 & 0.8681 & 0.7572 & 0.7159 & 0.4335 & 0.3055 \\
&& IQP & 61.0963 & 0.5665 & 0.9894 & 0.9894 & 0.7205 & 0.6058 & 0.2138 & 0.1044 \\
&& Displacement & 71.3904 & 0.7538 & 0.6464 & 0.6464 & 0.6960 & 0.7148 & 0.4288 & 1.3152 \\
&& Squeezing & 71.3904 & 0.7538 & 0.6464 & 0.6464 & 0.6960 & 0.7148 & 0.4288 & 0.1623 \\
\cline{2-11}
& \multirow{4}{*}{\makecell[{{p{1.0cm}}}]{15}}
& Classical & 72.1925 & 0.7031 & 0.7810 & 0.7810 & 0.7400 & 0.7211 & 0.4429 & 0.3926 \\
&& IQP & 60.4278 & 0.6456 & 0.4855 & 0.4855 & 0.5542 & 0.6059 & 0.2111 & 0.2815 \\
&& Displacement & 66.3102 & 0.6576 & 0.6992 & 0.6992 & 0.6777 & 0.6626 & 0.3255 & 18.6747 \\
&& Squeezing & 72.3262 & 0.7183 & 0.7467 & 0.7467 & 0.7322 & 0.7229 & 0.4461 & 0.6358 \\
\cline{2-11}
& \multirow{4}{*}{\makecell[{{p{1.0cm}}}]{23}}
& Classical & 74.0642 & 0.7273 & 0.7810 & 0.7810 & 0.7532 & 0.7401 & 0.4807 & 0.4581 \\
&& IQP & 60.4278 & 0.5719 & 0.8707 & 0.8707 & 0.6904 & 0.6007 & 0.2028 & 0.2929 \\
&& Displacement & 70.9893 & 0.6875 & 0.7836 & 0.7836 & 0.7324 & 0.7089 & 0.4186 & 42.9444 \\
&& Squeezing & 73.7968 & 0.7328 & 0.7599 & 0.7599 & 0.7461 & 0.7377 & 0.4756 & 0.5365 \\
\hline
\pagebreak
\multirow{12}{*}{\makecell[{{p{1.5cm}}}]{SVM RBF}}  &
\multirow{4}{*}{\makecell[{{p{1.0cm}}}]{2}} 
& Classical & 75.4011 & 0.7468 & 0.7784 & 0.7784 & 0.7623 & 0.7537 & 0.5077 & 0.3022 \\
&& IQP & 61.0963 & 0.5665 & 0.9894 & 0.9894 & 0.7205 & 0.6058 & 0.2138 & 0.1794 \\
&& Displacement & 71.3904 & 0.7538 & 0.6464 & 0.6464 & 0.6960 & 0.7148 & 0.4288 & 0.4493 \\
&& Sqeezing & 71.3904 & 0.7538 & 0.6464 & 0.6464 & 0.6960 & 0.7148 & 0.4288 & 0.2398 \\
\cline{2-11}
& \multirow{4}{*}{\makecell[{{p{1.0cm}}}]{15}}
& Classical & 73.6631 & 0.7241 & 0.7757 & 0.7757 & 0.7490 & 0.7361 & 0.4727 & 0.4317 \\
&& IQP & 64.3048 & 0.6600 & 0.6095 & 0.6095 & 0.6337 & 0.6435 & 0.2867 & 0.2848 \\
&& Displacement & 73.9305 & 0.7421 & 0.7441 & 0.7441 & 0.7431 & 0.7392 & 0.4785 & 0.5057 \\
&& Sqeezing & 73.9305 & 0.7421 & 0.7441 & 0.7441 & 0.7431 & 0.7392 & 0.4785 & 0.4868 \\
\cline{2-11}
& \multirow{4}{*}{\makecell[{{p{1.0cm}}}]{23}}
& Classical & 77.2727 & 0.7700 & 0.7863 & 0.7863 & 0.7781 & 0.7725 & 0.5452 & 0.5184 \\
&& IQP & 63.6364 & 0.6126 & 0.7678 & 0.7678 & 0.6815 & 0.6346 & 0.2701 & 0.4064 \\
&& Displacement & 74.5989 & 0.7455 & 0.7573 & 0.7573 & 0.7513 & 0.7458 & 0.4918 & 0.4938 \\
&& Sqeezing & 74.5989 & 0.7455 & 0.7573 & 0.7573 & 0.7513 & 0.7458 & 0.4918 & 0.5166 \\
\hline
\multirow{12}{*}{\makecell[{{p{1.5cm}}}]{SVM \\ Sigmoid}}  & 
\multirow{4}{*}{\makecell[{{p{1.0cm}}}]{2}} 
&Classical & 67.3797 & 0.6870 & 0.6544 & 0.6544 & 0.6703 & 0.6741 & 0.3479 & 0.4282 \\
&& IQP & 61.0963 & 0.5665 & 0.9894 & 0.9894 & 0.7205 & 0.6058 & 0.2138 & 0.1040 \\
&& Displacement & 49.3316 & 0.0000 & 0.0000 & 0.0000 & 0.0000 & 0.5000 & 0.0000 & 0.5616 \\
&& Sqeezing & 62.0321 & 0.6179 & 0.6570 & 0.6570 & 0.6368 & 0.6198 & 0.2399 & 0.3813 \\
\cline{2-11}
& \multirow{4}{*}{\makecell[{{p{1.0cm}}}]{15}} 
& Classical & 71.3904 & 0.7037 & 0.7520 & 0.7520 & 0.7270 & 0.7134 & 0.4272 & 0.4887 \\
&& IQP & 57.0856 & 0.5780 & 0.5673 & 0.5673 & 0.5726 & 0.5709 & 0.1418 & 0.1729 \\
&& Displacement & 49.3316 & 0.0000 & 0.0000 & 0.0000 & 0.0000 & 0.5000 & 0.0000 & 0.4445 \\
&& Sqeezing & 60.5615 & 0.6123 & 0.6042 & 0.6042 & 0.6082 & 0.6056 & 0.2112 & 0.4776 \\
\cline{2-11}
& \multirow{4}{*}{\makecell[{{p{1.0cm}}}]{23}}
&Classical & 74.0642 & 0.7330 & 0.7678 & 0.7678 & 0.7500 & 0.7403 & 0.4809 & 0.5157 \\
&& IQP & 52.1390 & 0.5279 & 0.5251 & 0.5251 & 0.5265 & 0.5213 & 0.0427 & 0.2880 \\
&& Displacement & 49.3316 & 0.0000 & 0.0000 & 0.0000 & 0.0000 & 0.5000 & 0.0000 & 0.4181 \\
&& Sqeezing & 55.4813 & 0.5599 & 0.5673 & 0.5673 & 0.5636 & 0.5546 & 0.1093 & 0.5428 \\
\hline
\pagebreak
\multirow{12}{*}{\makecell[{{p{1.5cm}}}]{Decision Tree}} &
\multirow{4}{*}{\makecell[{{p{1.5cm}}}]{2}}
& Classical & 66.7112 & 0.6757 & 0.6596 & 0.6596 & 0.6676 & 0.6672 & 0.3343 & 0.0096 \\
&& IQP & 61.0963 & 0.5665 & 0.9894 & 0.9894 & 0.7205 & 0.6058 & 0.2138 & 0.0010 \\
&& Displacement & 71.3904 & 0.7538 & 0.6464 & 0.6464 & 0.6960 & 0.7148 & 0.4288 & 0.00246 \\
&& Sqeezing & 71.3904 & 0.7538 & 0.6464 & 0.6464 & 0.6960 & 0.7148 & 0.4288 & 0.001596 \\
\cline{2-11}
& \multirow{4}{*}{\makecell[{{p{1.5cm}}}]{15}} 
& Classical & 64.8396 & 0.6457 & 0.6781 & 0.6781 & 0.6615 & 0.6480 & 0.2962 & 0.0629 \\
&& IQP & 64.1711 & 0.6521 & 0.6280 & 0.6280 & 0.6398 & 0.6419 & 0.2837 & 0.030999 \\
&& Displacement & 67.3797 & 0.6819 & 0.6675 & 0.6675 & 0.6747 & 0.6739 & 0.3477 & 0.007659 \\
&& Sqeezing & 66.8449 & 0.6747 & 0.6675 & 0.6675 & 0.6711 & 0.6685 & 0.3369 & 0.015341 \\
\cline{2-11}
& \multirow{4}{*}{\makecell[{{p{1.5cm}}}]{23}}
& Classical & 66.7112 & 0.6729 & 0.6675 & 0.6675 & 0.6702 & 0.6671 & 0.3342 & 0.1061 \\
&& IQP & 62.0321 & 0.6294 & 0.6095 & 0.6095 & 0.6193 & 0.6205 & 0.2408 & 0.1180 \\
&& Displacement & 68.5829 & 0.7057 & 0.6517 & 0.6517 & 0.6776 & 0.6863 & 0.3722 & 0.01805 \\
&& Sqeezing & 68.3155 & 0.6994 & 0.6570 & 0.6570 & 0.6776 & 0.6835 & 0.3667 & 0.011008  \\
\hline
\multirow{12}{*}{\makecell[{{p{1.5cm}}}]{Random Forest}} & 
\multirow{4}{*}{\makecell[{{p{1.5cm}}}]{2}}
& Classical & 72.5936 & 0.7302 & 0.7282 & 0.7282 & 0.7292 & 0.7259 & 0.4518 & 0.5493 \\
&& IQP & 61.0963 & 0.5665 & 0.9894 & 0.9894 & 0.7205 & 0.6058 & 0.2138 & 0.0952 \\
&& Displacement & 71.3904 & 0.7538 & 0.6464 & 0.6464 & 0.6960 & 0.7148 & 0.4288 & 0.3154 \\
&& Sqeezing & 71.3904 & 0.7538 & 0.6464 & 0.6464 & 0.6960 & 0.7148 & 0.4288 & 0.3261 \\
\cline{2-11}
&\multirow{4}{*}{\makecell[{{p{1.5cm}}}]{15}} 
& Classical & 74.1979 & 0.7337 & 0.7704 & 0.7704 & 0.7516 & 0.7416 & 0.4835 & 1.0324 \\
&& IQP & 69.1176 & 0.7056 & 0.6702 & 0.6702 & 0.6874 & 0.6915 & 0.3827 & 0.5762 \\
&& Displacement & 71.3904 & 0.7110 & 0.7335 & 0.7335 & 0.7221 & 0.7136 & 0.4275 & 0.5088 \\
&& Sqeezing & 70.4545 & 0.7047 & 0.7177 & 0.7177 & 0.7111 & 0.7044 & 0.4088 & 0.4091\\
\cline{2-11}
& \multirow{4}{*}{\makecell[{{p{1.5cm}}}]{23}} 
& Classical & 75.4011 & 0.7701 & 0.7335 & 0.7335 & 0.7514 & 0.7543 & 0.5082 & 1.3224 \\
&& IQP & 65.6417 & 0.6614 & 0.6596 & 0.6596 & 0.6605 & 0.6564 & 0.3127 & 1.2142 \\
&& Displacement & 71.3904 & 0.7165 & 0.7203 & 0.7203 & 0.7184 & 0.7138 & 0.4277 & 0.4622 \\
&& Sqeezing & 71.5241 & 0.7184 & 0.7203 & 0.7203 & 0.7194 & 0.7152 & 0.4304 & 0.4479 \\
\hline		
\pagebreak	
\multirow{12}{*}{\makecell[{{p{1.5cm}}}]{LightGBM}} & 
\multirow{4}{*}{\makecell[{{p{1.5cm}}}]{2}} 
&Classical & 73.9305 & 0.7359 & 0.7573 & 0.7573 & 0.7464 & 0.7391 & 0.4783 & 0.1178 \\
&&IQP & 61.0963 & 0.5665 & 0.9894 & 0.9894 & 0.7205 & 0.6058 & 0.2138 & 0.0127 \\
&&Displacement & 71.3904 & 0.7538 & 0.6464 & 0.6464 & 0.6960 & 0.7148 & 0.4288 & 0.0978 \\
&&Sqeezing & 71.3904 & 0.7538 & 0.6464 & 0.6464 & 0.6960 & 0.7148 & 0.4288 & 0.0261 \\
\cline{2-11}
& \multirow{4}{*}{\makecell[{{p{1.5cm}}}]{15}}
&Classical & 74.7326 & 0.7230 & 0.8127 & 0.8127 & 0.7652 & 0.7464 & 0.4937 & 0.2598 \\
&&IQP & 67.2460 & 0.6831 & 0.6596 & 0.6596 & 0.6711 & 0.6726 & 0.3451 & 0.0719 \\
&&Displacement & 71.6578 & 0.7093 & 0.7467 & 0.7467 & 0.7275 & 0.7162 & 0.4326 & 0.0838 \\
&&Sqeezing & 71.6578 & 0.7093 & 0.7467 & 0.7467 & 0.7275 & 0.7162 & 0.4326 & 0.0867 \\
\cline{2-11}
& \multirow{4}{*}{\makecell[{{p{1.5cm}}}]{23}} 
&Classical & 76.6043 & 0.7727 & 0.7625 & 0.7625 & 0.7676 & 0.7661 & 0.5321 & 0.4129 \\
&&IQP & 65.2406 & 0.6639 & 0.6359 & 0.6359 & 0.6496 & 0.6526 & 0.3051 & 0.1141 \\
&&Displacement & 71.9251 & 0.7183 & 0.7335 & 0.7335 & 0.7258 & 0.7191 & 0.4382 & 0.0956 \\
&&Sqeezing & 71.9251 & 0.7183 & 0.7335 & 0.7335 & 0.7258 & 0.7191 & 0.4382 & 0.1211 \\
\hline
\multirow{12}{*}{\makecell[{{p{1.5cm}}}]{AdaBoost}} & 
\multirow{4}{*}{\makecell[{{p{1.5cm}}}]{2}} 
& Classical & 75.2674 & 0.7450 & 0.7784 & 0.7784 & 0.7613 & 0.7523 & 0.5050 & 0.3176 \\
&& IQP & 61.0963 & 0.5665 & 0.9894 & 0.9894 & 0.7205 & 0.6058 & 0.2138 & 0.0608 \\
&& Displacement & 71.3904 & 0.7538 & 0.6464 & 0.6464 & 0.6960 & 0.7148 & 0.4288 & 0.1183 \\
&& Sqeezing & 71.3904 & 0.7538 & 0.6464 & 0.6464 & 0.6960 & 0.7148 & 0.4288 & 0.1330 \\
\cline{2-11}
& \multirow{4}{*}{\makecell[{{p{1.5cm}}}]{15}} 
& Classical & 74.5989 & 0.7405 & 0.7678 & 0.7678 & 0.7539 & 0.7457 & 0.4916 & 0.4682 \\
&& IQP & 64.8396 & 0.6495 & 0.6649 & 0.6649 & 0.6571 & 0.6482 & 0.2964 & 0.2187 \\
&& Displacement & 71.5241 & 0.7318 & 0.6913 & 0.6913 & 0.7110 & 0.7156 & 0.4308 & 0.1559 \\
&& Sqeezing & 71.5241 & 0.7318 & 0.6913 & 0.6913 & 0.7110 & 0.7156 & 0.4308 & 0.1765 \\
\cline{2-11}
& \multirow{4}{*}{\makecell[{{p{1.5cm}}}]{23}} 
& Classical & 77.5401 & 0.7726 & 0.7889 & 0.7889 & 0.7807 & 0.7752 & 0.5506 & 0.6542 \\
&& IQP & 67.2460 & 0.6516 & 0.7599 & 0.7599 & 0.7016 & 0.6713 & 0.3433 & 0.4047 \\
&& Displacement & 72.8610 & 0.7316 & 0.7335 & 0.7335 & 0.7325 & 0.7285 & 0.4571 & 0.1799 \\
&& Sqeezing & 72.8610 & 0.7316 & 0.7335 & 0.7335 & 0.7325 & 0.7285 & 0.4571 & 0.1830 \\
\hline
\pagebreak
\multirow{12}{*}{\makecell[{{p{1.5cm}}}]{CatBoost}} &
 \multirow{4}{*}{\makecell[{{p{1.5cm}}}]{2}} 
& Classical & 74.7326 & 0.7375 & 0.7784 & 0.7784 & 0.7574 & 0.7469 & 0.4942 & 5.5071 \\
&& IQP & 61.0963 & 0.5665 & 0.9894 & 0.9894 & 0.7205 & 0.6058 & 0.2138 & 1.3892 \\
&& Displacement & 71.3904 & 0.7538 & 0.6464 & 0.6464 & 0.6960 & 0.7148 & 0.4288 & 0.9829 \\
&& Sqeezing & 71.3904 & 0.7538 & 0.6464 & 0.6464 & 0.6960 & 0.7148 & 0.4288 & 0.9947 \\
\cline{2-11}
& \multirow{4}{*}{\makecell[{{p{1.5cm}}}]{15}} 
& Classical & 74.8663 & 0.7335 & 0.7916 & 0.7916 & 0.7614 & 0.7481 & 0.4967 & 18.9151 \\
&& IQP & 68.4492 & 0.6948 & 0.6728 & 0.6728 & 0.6836 & 0.6847 & 0.3691 & 3.1672 \\
&& Displacement & 71.1230 & 0.7032 & 0.7441 & 0.7441 & 0.7231 & 0.7108 & 0.4219 & 3.5555 \\
&& Sqeezing & 71.1230 & 0.7032 & 0.7441 & 0.7441 & 0.7231 & 0.7108 & 0.4219 & 3.0286 \\
\cline{2-11}
& \multirow{4}{*}{\makecell[{{p{1.5cm}}}]{23}} 
& Classical & 77.6738 & 0.7790 & 0.7810 & 0.7810 & 0.7800 & 0.7767 & 0.5534 & 11.7024 \\
&& IQP & 67.2460 & 0.6754 & 0.6807 & 0.6807 & 0.6781 & 0.6723 & 0.3447 & 5.6349 \\
&& Displacement & 73.1283 & 0.7330 & 0.7388 & 0.7388 & 0.7359 & 0.7312 & 0.4624 & 2.4728 \\
&& Sqeezing & 73.1283 & 0.7330 & 0.7388 & 0.7388 & 0.7359 & 0.7312 & 0.4624 & 2.4147 \\
\hline
\multirow{12}{*}{\makecell[{{p{1.5cm}}}]{Extra Trees}} & 
\multirow{4}{*}{\makecell[{{p{1.5cm}}}]{2}}
& Classical & 72.5936 & 0.7186 & 0.7546 & 0.7546 & 0.7362 & 0.7255 & 0.4514 & 0.4512 \\
&& IQP & 61.0963 & 0.5665 & 0.9894 & 0.9894 & 0.7205 & 0.6058 & 0.2138 & 0.0764 \\
&& Displacement & 71.3904 & 0.7538 & 0.6464 & 0.6464 & 0.6960 & 0.7148 & 0.4288 & 0.1328 \\
&& Sqeezing & 71.3904 & 0.7538 & 0.6464 & 0.6464 & 0.6960 & 0.7148 & 0.4288 & 0.1466 \\
\cline{2-11}
& \multirow{4}{*}{\makecell[{{p{1.5cm}}}]{15}} 
& Classical & 70.9893 & 0.7035 & 0.7388 & 0.7388 & 0.7207 & 0.7095 & 0.4193 & 0.4377 \\
&& IQP & 67.7807 & 0.6983 & 0.6412 & 0.6412 & 0.6685 & 0.6783 & 0.3562 & 0.2745 \\
&& Displacement & 69.2513 & 0.6966 & 0.6966 & 0.6966 & 0.6966 & 0.6925 & 0.3849 & 0.4256 \\
&& Sqeezing & 68.9840 & 0.6960 & 0.6887 & 0.6887 & 0.6923 & 0.6899 & 0.3797 & 0.4381 \\
\cline{2-11}
& \multirow{4}{*}{\makecell[{{p{1.5cm}}}]{23}} 
& Classical & 74.1979 & 0.7541 & 0.7282 & 0.7282 & 0.7409 & 0.7422 & 0.4841 & 0.4781 \\
&& IQP & 64.9733 & 0.6552 & 0.6517 & 0.6517 & 0.6534 & 0.6497 & 0.2994 & 0.3730 \\
&& Displacement & 72.1925 & 0.7355 & 0.7045 & 0.7045 & 0.7197 & 0.7222 & 0.4441 & 0.4927 \\
&& Sqeezing & 72.7273 & 0.7384 & 0.7150 & 0.7150 & 0.7265 & 0.7274 & 0.4547 & 0.4869 \\
\hline
\pagebreak
\multirow{12}{*}{\makecell[{{p{1.5cm}}}]{Gradient  Boosting}}
&\multirow{4}{*}{\makecell[{{p{1.5cm}}}]{2}} 
& Classical & 74.8663 & 0.7370 & 0.7836 & 0.7836 & 0.7596 & 0.7482 & 0.4968 & 0.6142 \\
&& IQP & 61.0963 & 0.5665 & 0.9894 & 0.9894 & 0.7205 & 0.6058 & 0.2138 & 0.0653 \\
&& Displacement & 71.3904 & 0.7538 & 0.6464 & 0.6464 & 0.6960 & 0.7148 & 0.4288 & 0.1589 \\
&& Squeezing & 71.3904 & 0.7538 & 0.6464 & 0.6464 & 0.6960 & 0.7148 & 0.4288 & 0.1425 \\
\cline{2-11}
&\multirow{4}{*}{\makecell[{{p{1.5cm}}}]{15}} 
& Classical & 74.3316 & 0.7221 & 0.8021 & 0.8021 & 0.7600 & 0.7425 & 0.4858 & 1.7732 \\
&& IQP & 68.8503 & 0.6995 & 0.6755 & 0.6755 & 0.6872 & 0.6887 & 0.3772 & 0.8825 \\
&& Displacement & 73.7968 & 0.7305 & 0.7652 & 0.7652 & 0.7474 & 0.7376 & 0.4755 & 0.3557 \\
&& Squeezing & 73.7968 & 0.7305 & 0.7652 & 0.7652 & 0.7474 & 0.7376 & 0.4755 & 0.347 \\
\cline{2-11}
& \multirow{4}{*}{\makecell[{{p{1.5cm}}}]{23}} 
& Classical & 78.4759 & 0.7884 & 0.7863 & 0.7863 & 0.7873 & 0.7847 & 0.5695 & 2.6064 \\
&& IQP & 66.1765 & 0.6591 & 0.6887 & 0.6887 & 0.6735 & 0.6614 & 0.3230 & 2.0365 \\
&& Displacement & 74.8663 & 0.7468 & 0.7625 & 0.7625 & 0.7546 & 0.7485 & 0.4971 & 0.4710 \\
&& Sqeezing& 74.8663 & 0.7468 & 0.7625 & 0.7625 & 0.7546 & 0.7485 & 0.4971 & 0.4179 \\
\hline
\multirow{12}{*}{\makecell[{{p{1.5cm}}}]{XGBoost}}
& \multirow{4}{*}{\makecell[{{p{1.5cm}}}]{2}}
& Classical & 74.3316 & 0.7379 & 0.7652 & 0.7652 & 0.7513 & 0.7430 & 0.4863 & 0.6346 \\
&& IQP & 61.0963 & 0.5665 & 0.9894 & 0.9894 & 0.7205 & 0.6058 & 0.2138 & 0.0393 \\
&& Displacement & 71.3904 & 0.7538 & 0.6464 & 0.6464 & 0.6960 & 0.7148 & 0.4288 & 0.1746 \\
&& Squeezing & 71.3904 & 0.7538 & 0.6464 & 0.6464 & 0.6960 & 0.7148 & 0.4288 & 0.0263\\
\cline{2-11}
& \multirow{4}{*}{\makecell[{{p{1.5cm}}}]{15}}
& Classical & 73.3957 & 0.7239 & 0.7678 & 0.7678 & 0.7452 & 0.7335 & 0.4674 & 8.6842 \\
&& IQP & 66.7112 & 0.6711 & 0.6728 & 0.6728 & 0.6719 & 0.6670 & 0.3341 & 0.0897 \\
&& Displacement & 68.8503 & 0.6807 & 0.7256 & 0.7256 & 0.7024 & 0.6880 & 0.3763 & 0.0787 \\
&& Squeezing & 68.8503 & 0.6807 & 0.7256 & 0.7256 & 0.7024 & 0.6880 & 0.3763 & 0.0810 \\
\cline{2-11}
& \multirow{4}{*}{\makecell[{{p{1.5cm}}}]{23}} 
& Classical & 74.0642 & 0.7493 & 0.7335 & 0.7335 & 0.7413 & 0.7407 & 0.4813 & 5.7666 \\
&& IQP & 63.5027 & 0.6417 & 0.6332 & 0.6332 & 0.6375 & 0.6351 & 0.2701 & 0.2480 \\
&& Displacement & 70.3209 & 0.7199 & 0.6781 & 0.6781 & 0.6984 & 0.7035 & 0.4068 & 0.0953 \\
&& Squeezing & 70.3209 & 0.7199 & 0.6781 & 0.6781 & 0.6984 & 0.7035 & 0.4068 & 0.1266 \\

\hline
\end{longtable}
\end{landscape}

\section{Conclusion}

This study explored the impact of different data encoding techniques, including Classical, IQP, Displacement, and Squeezing, on the performance of various machine learning models across multiple PCA dimensions. The results indicate that encoding strategies significantly influence classification accuracy, computational efficiency, and model behavior. PCA dimensionality reduction consistently improves model performance, with PCA=23 yielding the highest accuracy across most scenarios. However, the choice of encoding plays a crucial role, as IQP encoding consistently underperforms, while Displacement and Squeezing encodings demonstrate notable improvements. \\ Despite their potential, quantum encoding techniques introduce computational overhead, particularly in high-dimensional feature spaces. While certain quantum encodings improve classification accuracy, their feasibility in practical applications depends on optimizing computational efficiency. The findings emphasize the need for further investigation into hybrid quantum-classical models to balance accuracy and efficiency. Future research should focus on refining encoding mechanisms, exploring alternative feature selection techniques, and leveraging quantum advantage in real-world predictive modeling tasks.
\section{Conflict of Interest Statement}
The authors declare that there is no conflict of interest regarding the publication of this manuscript.

\section*{Declarations}

\subsection*{Ethical Approval and Consent to Participate}
Not applicable. This study did not involve any human or animal subjects requiring ethical approval.

\subsection*{Consent for Publication}
Not applicable. This manuscript does not contain any individual person’s data in any form.

\subsection*{Availability of Supporting Data}
The data that support the findings of this study are available from the corresponding author upon reasonable request.

\subsection*{Competing Interests/Authors' Contributions}
The authors declare that they have no competing interests.

\textbf{Authors' Contributions:}
All authors contributed equally.

\subsection*{Funding}
This research received no specific grant from any funding agency in the public, commercial, or not-for-profit sectors.

\bibliographystyle{unsrt}
\bibliography{AdvancedQDataEncoding}
	
\section*{Data Availability Statement}
The datasets analysed during the current study are available in the keggal repository, https://www.kaggle.com/datasets/blastchar/telco-customer-churn

\section*{Author Information}

Authors and Affiliations:\\
\textbf{PhD of Analytics and Decision Science}, IIM Mumbai, India.\\
\href{mailto:minati.rath.2019@iimmumbai.co.in}{Minati Rath}.\\
\textbf{Professor of Analytics and Decision Science}, IIM Mumbai, India.\\
\href{mailto:hemadate@iimmumbai.ac.in}{Hema Date}.

\section*{Corresponding Authors}
Correspondence to \href{mailto:minati.rath.2019@iimmumbai.ac.in}{Minati Rath} or \href{mailto:hemadate@iimmumbai.ac.in}{Hema Date}.

\end{document}